# Disability data futures:
# Achievable imaginaries for AI and disability data justice

Denis Newman-Griffis[1], Bonnielin Swenor[2], Rupa Valdez[3], Gillian Mason[4]

[1] Information School, University of Sheffield, Sheffield, UK
[2] Disability Health Research Center, Johns Hopkins University, Baltimore, MD, USA
[3] School of Engineering and Applied Sciences, and School of Medicine, University of Virginia, Charlottesville, VA, USA
[4] Independent disability advocate and clinician-researcher, NSW, Australia

## Abstract

Data are the medium through which individuals' identities and experiences are filtered in contemporary states and systems, and AI is increasingly the layer mediating between people, data, and decisions. The history of data and AI is often one of disability exclusion, oppression, and the reduction of disabled experience; left unchallenged, the current proliferation of AI and data systems thus risks further automating ableism behind the veneer of algorithmic neutrality. However, exclusionary histories do not preclude inclusive futures, and disability-led visions can chart new paths for collective action to achieve futures founded in disability justice. This chapter brings together four academics and disability advocates working at the nexus of disability, data, and AI, to describe *achievable imaginaries* for artificial intelligence and disability data justice. Reflecting diverse contexts, disciplinary perspectives, and personal experiences, we draw out the shape, actors, and goals of imagined future systems where data and AI support movement towards disability justice.

## Introduction

To be a human being within contemporary states and systems is to be a data subject. The complexities of human experience, and competing ideas and understandings of well-being, are increasingly materialised through ever-growing sources of data. As reliance on these data grows, they transcend being mere representations and become the primary source of truth in understanding what it looks like to be human and to be well (Oman, 2021). Data therefore have significant power, and exclusion from or misrepresentation in data results in exclusion and harm from the processes that use those data (Dalmer et al., 2024).

Disabled people frequently experience just such exclusion and misrepresentation in data systems. We focus on *disability data*, described by Aboulafia et al. (2024) as data 'collected and used to quantify and generate insights about people with disabilities,' and including also broader data regarding a person's identity and experience of disability (Newman-Griffis et al., 2022).

Disability data are contested and deeply political, and affected by structural ableism throughout policy and health systems (Valdez and Swenor, 2023). National and population surveys use competing definitions and measurement strategies, making their data a source of fierce debate and meaning that millions of disabled people quite literally do not 'count' (Hahn, 1993; Swenor and Landes, 2023). For individuals, conceptual, practical, and technological barriers lead to loss of disability data and create 'information inequities' in health and care data (Morris et al., 2022; Newman-Griffis et al., 2022). These data harms, in which vital aspects of experience, identity, and priorities are omitted or lost from data and thus from care systems, are often magnified by the use of artificial intelligence methodologies to process and analyse disability data. The design of AI systems can further embed oppressive epistemologies of disability, such as medicalised definitions and exclusionary policies, whilst operating under the guise of algorithmic neutrality (Newman-Griffis et al., 2023; Weerts et al., 2024).

However, disability injustice is not a foregone conclusion. Engaging with disability communities in rethinking disability data and data systems can help to foster disability data justice (Aboulafia et al., 2024; Swenor, 2022; Valdez et al., 2022). Designing new data and AI systems that are sensitive to disability data challenges can help to overcome information barriers and develop more person-centred data (Newman-Griffis et al., 2022). As Thatcher and Dalton argue for spatial data (Thatcher and Dalton, 2022), we can reimagine approaches to disability-led data and data systems to help challenge structures of oppression and explore new, emancipatory data and technologies for disability liberation.

This chapter presents visions of emancipatory disability data futures to move us towards disability data justice and AI justice. We are four scholars, practitioners, and advocates working at the intersection of disability, data, and AI. We have all experienced and critiqued the limitations and the harms of current disability data and AI, but we embrace the potential of transformative futures to challenge the status quo and chart new paths for collective action on data and AI.

We approach these futures through the lens of *social imaginaries* (Taylor, 2002), a strategy for exploring alternative social existences and the behaviours and collective notions that underpin them. Social imaginaries have been used to explore and define new, desirable futures in data and AI (Lehtiniemi and Ruckenstein, 2019; Rahm and Rahm-Skågeby, 2023; Sartori and Bocca, 2023). Imaginaries provide a valuable tool for imagining new data systems free from—and/or freely responding to—the constraints of current practice.

To bring our perspectives together, we adopted the method of collective writing, a pioneering approach designed to engage with a multiplicity of perspectives and voices to reflect the 'continuous struggle for meaning-making' (Jandrić et al., 2023). The 'diffractive writing' process this creates (Jarke and Bates, 2024), in which the harmonies and tensions between different author contributions reflect the dynamic nature of collaborative research, aligns for us with the multiplicity and community-driven nature of disability discourses.

## Reimagining records to reshape AI - *Denis Newman-Griffis*

*Denis is a white, gender/queer, neurodivergent academic from a middle-class background. Their work is situated in traditions of theory and practice from AI and data science, health informatics, and disability studies.*

Artificial intelligence is both a prism and a megaphone. The use of AI refracts the many characters and histories embedded in data into discrete, filtered categories, while amplifying the voices and perspectives that produced the data (Crawford and Paglen, 2021). Troubling the role of AI in realising disability justice is in part a matter of troubling the data on which AI depends and which it transforms. Few data are more troublesome of, and to, disability than the medical record.

The medical record has long been a site of plural and competing epistemologies (Greenhalgh et al., 2009). On one level a record of fact to support the provision of timely, appropriate care, medical records also shape the practice of care itself, determining how care is enacted and sustained (Halford et al., 2010; Isah and Byström, 2020). The design and implementation of medical record systems is governed by financial pressures to account for time and services, yet medical records become authoritative sources of medical knowledge (Koopman et al., 2022), and produce the single, stable truth of a patient's history (Berg and Bowker, 1997). Operating within this record, AI systems reify the infrastructure of care and automate its efficiencies, reinforcing the production of sole medical truth by care professionals.

In contrast to the external, organising record, we imagine an alternative, multilateral health record co-created for dialogue and collaboration. Rather than a stable and temporally-unfolding history of fact, our imagined record operates as a dynamic space for reflecting and collaboratively shaping the fundamentally unstable nature of disabled experience. This is a record of plural truths and lived multiplicity, including data that reflect an intertwined mixture of experiences, perspectives, priorities, and histories from not only patient and provider, but also wider networks of interdependence. This collective and dynamic record thus materialises a more collaborative process and infrastructure of care, focused on collective response to lived experience and co-produced action towards disabled people's personal priorities.

Acting in this record, AI systems mediate through multiplicity: they are designed to help organise and bring together diverse perspectives and information from the record, and to maintain the tensions between them rather than collapsing to a single, dominating—and synthetic—truth. Simultaneously, AI systems also facilitate the co-creation and maintenance of the record, through intentional interaction and recording (such as speech recognition and assisted writing) and ambient engagement with contextual information to situate recorded data (through channels such as wearable devices and sensors). AI thus both mediates and facilitates creating, accessing, and acting on the record, doing so in all directions within a collaborative, agentic framework rather than unidirectionally imposing process and practice upon a passively receiving patient.

This new health record is not too distant to reach. The growth of personal health records, dynamic and evolving through multiple authorship and visibility in patient portals (Hawthorne and Richards, 2017), is laying the groundwork for a health record that can speak and advocate with multiple histories. AI-assisted recording systems are becoming more commonplace, reducing barriers to editing access to the record (Blackley et al., 2019; Lu et al., 2020). Turning these facilitators into tangible change requires taking a co-productive and disability-led approach to redesigning the purposes and processes that the record implements. Doing so will reshape the nature of medical history, and reimagine the AI systems that act in and are shaped by the multiple stories of the health record.

## Improving disability data is foundational to AI - *Bonnielin Swenor*

*Bonnie is a white disabled woman whose research focuses on using data to advance equity and justice for people with disabilities.*

To identify and address AI bias impacting disabled people, we first need data to identify this population (Aboulafia et al., 2024). However, unaddressed and underlying data limitations prevent us from determining where and when AI disability bias occurs.

Current methods used to collect data on disability or disability status have been shown to underestimate disability prevalence. These measures miss many groups of disabled people, including but not limited to people with long COVID and other chronic conditions, as well as people with psychiatric disabilities (Hall et al., 2022; Landes et al., 2024b; Schulz et al., 2024). Contemporary measures of disability primarily focus on assessing an individual's functioning, a separate but overlapping construct. However, not every disabled person experiences functional limitations (Hall et al., 2022), and disabled people do not need to experience functional limitations to be covered by disability civil rights protections in some countries, such as the United States (Landes et al., 2024a). As a result, current measures of disability misclassify disabled people as not having a disability, resulting in inaccurate estimates of disability prevalence. These data limitations also make it challenging to identify disabled people within datasets, including disabled people from intersecting marginalised groups, limiting our ability to identify and address algorithmic and AI biases impacting this population.

Additionally, disabled people are often excluded from datasets due to unaddressed barriers in the data collection process. For example, recent data indicate that out of thirty U.S. federal, nationally representative surveys, none had protocols to make data collection accessible and inclusive of disabled people (Cerilli et al., 2024). These data underscore how disabled people are excluded from datasets at every stage and level. Yet, few strategies or toolkits focus on ensuring disabled people are included in data collection efforts. The exclusion of disabled people at the data collection level calls into question the generalizability and representativeness of datasets used to develop, test, and inform AI.

Relatedly, the inaccessibility of data itself is problematic. Barriers to accessing, using, and understanding data exclude disabled people from accessing critical information, which, in turn, limits opportunities for disabled communities to identify and call out biases in data. In part, this stems from the historic exclusion of disabled people from data science. Yet, accessibility is everyone's responsibility. The barriers to data access stem from the lack of knowledge about accessibility and universal design across data and AI professions.

Disability data justice principles (Swenor, 2022) can be used to redress the harms outlined above and help maximise the potential of AI. Such an approach focuses on developing strategies to address the historical exclusion of disabled people from data collection, analysis, and dissemination. The foundation of disability data justice is ensuring disabled people are part of all parts and phases of data-driven research and approaches, including

AI. However, disability data justice should not be limited to efforts focused on disabled people or their data alone. Instead, there must be recognition that all issues are disability issues, and all AI efforts must work to uphold and advance disability data justice.

## Synthesising cross-disciplinary methods for disability-inclusive AI - *Rupa Valdez*

*Rupa is a chronically ill and disabled South Asian woman from a middle-class background who lives in the United States. Her work lies at the intersection of human factors engineering, health informatics, cultural anthropology, and community engagement.*

Addressing the needs of the disability community in an era of AI requires approaches that centre the community across problem definition, design implementation, and evaluation and encourage continuous improvement through a commitment to an iterative process. One way to reimagine the ways in which AI is created and used is to synthesise the principles and techniques of user centred design, human centred design, and community-based participatory research.

Stemming from the disability rights movement, *universal design (UD)* is an approach to design products, environments, and systems that are "usable by all people, to the greatest extent possible, while minimising the need for adaptation or specialised design" (North Carolina State University College of Design, n.d.). UD asks the designer to be guided by seven overarching principles: equitable use, flexible use, simple and intuitive use, perceptible information, tolerance for error, low physical effort, and size and space for approach and use. Ideally, these principles guide the design effort from the very beginning; but may also be used to invite adaptation of tools and systems to expand usability. With its focus on use, universal design gives us particular insight into ways to make interfaces to engage with AI more accessible to all. Extending universal access to conversations about AI also leads us to reimagining ways in which AI may be explainable to all in ways that make creative use of visualisations, plain language, and multi-modal resources.

While universal design is a principle-driven design approach, *human-centred design (HCD)* is driven by engagement. HCD is guided by the belief that improving the usefulness and usability of products and services requires the active participation of its ultimate end users. As such, HCD requires the designer to interact with users throughout all stages of the design process. Moreover, it requires designers to see such involvement as an iterative process in which feedback from users may require revisiting earlier stages of the design process.

Engagement must explicitly include people who identify across physical, sensory, cognitive, and mental health disabilities and with those who hold a wide range of intersecting identities. In addition to disabled people themselves, those in their social and care networks and community-based organisations with a focus on disability rights and disability justice should be involved. Techniques such as maximum variance sampling may be used to facilitate such inclusion (Benda et al., 2020). Approaches to engagement are ideally longitudinal and move toward participatory design or co-design (Jolliff et al., 2023), but may begin with shorter-term engagements in the form of interviews, focus groups, surveys, and design workshops (Nadarzynski et al., 2024; Ozkaynak et al., 2021; Willis et al., 2023). Community engagement studios (Joosten et al., 2015), community advisory boards, and inclusion of disabled people on the design team also move the process toward co-production.

These latter approaches move us from what is traditionally within the scope of HCD toward approaches grounded in *community based participatory research (CBPR)*, a collaborative approach that emphasises the participation and influence of nonacademic researchers in all aspects of the process of creating knowledge (Israel et al., 1998). In addition to philosophically centring community voices, CBPR pushes traditional approaches within universal design and HCD towards creating AI systems that are usable and useful (including systems that do not perpetuate disparities) for disabled people, but also toward reimagining, from a disability community perspective, which types of problems AI should be used to address.

Although much work remains to be done to formally synthesise such approaches for AI, an early approach to such synthesis, grounded in lessons learned from working with the disability community, is articulated by Valdez et al. (2022).

## Shaping intersectional disability justice systems through leadership and culture - *Gillian Mason*

*Gillian is a white, queer and proudly disabled clinician-researcher and patient representative who lives on the unceded land of the Awabakal and Worimi people, in Australia. She draws on her background in rehabilitation, digital health, health technology assessment and governance.*

What if the designers, policy-makers and leaders in our data and AI ecosystems were disabled? A reimagined future in which people with disabilities are represented in the leadership and governance structures of the institutions developing, implementing, overseeing and regulating our systems disrupts the preconception that inclusive policy and practice is solely for end-users, participants or patients as passive recipients of support. This thinking is situated in a medical, charity or welfare model of disability (United Nations Department of Economic and Social Affairs, 2022). The United Nations Convention on the Rights of Persons with Disabilities insists that a human-rights model, where disability is embraced as a valued part of human diversity for all, is key to inclusive development.

Whilst diversity, equity and inclusion policies and strategies are increasingly recognised as crucial for the realisation of high-performing institutions that deliver impact (McKinsey & Company, 2023), not all diversity is equally supported, valued or leveraged. Disability is still often left out of diversity, equity and inclusion (DEI) initiatives (Wright, 2022), and when it is addressed it is often in terms of meeting minimum access requirements only, rather than pursuing true disability inclusion. In the absence of organisational support, disabled leaders themselves are taking on much of the responsibility and burden to create access and opportunities for themselves as well as supporting next generations of disabled leaders (Disability Leadership Institute, 2017; Orkin, 2023).

Organisations that respond to these needs will have three key characteristics. First, they adopt strengths-based, disability-inclusive mindsets, learning from cultural models of disability and inclusion that seek 'to improve the human condition through focussing on what keeps people strong, as distinct to merely negating the adverse impact of difference' (First People's Disability Network Australia, 2024). Second, they uplift capability by putting their policies into real practice, and adopting a continuous quality improvement mindset (Klinksiek, 2024; Singh and Meeks, 2023). Third, they intentionally cultivate the conditions for disability justice, including organisational culture, adapting co-design methodologies from research to share power and reshape organisational practice (Fraser-Barbour et al., 2023).

More than culture, disability-inclusive organisations and governments must allocate tangible resources. In academia, reducing disparities for disabled faculty requires resources to address ableism, create more inclusive environments, and raise standards beyond minimal regulatory compliance (Castro et al., 2024). Governments must lead by example, taking proactive action for inclusion beyond the blunt instrument of legislation, including shaping their own cultures and supporting broader cultural diversity and advocacy. There is much that can be done by amplifying existing disabled leaders and putting material support behind their work, and as Sara Rotenberg argues, doing so is necessary to avoid 'perpetuating inequity or stalling necessary progress' (Rotenberg, 2021).

Reimagining our focus on leadership and organisational culture through a disability justice lens will help to shape a future in which accessibility is understood as an act of love (Mingus, 2018), disability is a valued part of human diversity, and community-led participatory methods are standard practice for achieving inclusion and working towards justice.

## Concluding remarks

Achieving disability data justice, and building more just foundations for the intertwining of disability and AI, requires action on multiple fronts. The four visions presented in this chapter reflect the multiplicity of disability, data, and AI, and the range of disciplinary methods and stakeholders that must be involved in these actions.

Newman-Griffis illustrated the role of records in shaping disability data and AI, and envisioned new, disability-led records that provide the foundation for disability data justice. Swenor described current challenges in the definition and measurement of disability data, and identified actionable barriers to more just disability data. Valdez outlined a synthesis-based approach to bring together multiple design and research traditions in community-based, disability-centred approaches for taking action on disability data and AI. Finally, Mason highlighted the role of governments and institutions in bringing these ideas into tangible action.

Each of our visions draws on different traditions of research and advocacy, and each envisions different aspects of more inclusive disability data futures. In this multiplicity lies strength: just as the futures we strive for are plural and interdependent, so must our methods for achieving those futures weave together interdisciplinary methods and complementary perspectives.

Disability data futures are not fixed. Our imaginaries lay possible paths to reshaping these futures: to achieve more just data, and to co-create more just AI built on disability-led data foundations. These must be collective actions that work across traditional boundaries of discipline, experience, sector, and affiliation, and cannot solely be an academic exercise. The reward of taking these actions will be a more just future where data and AI work together to support disability justice.